\documentstyle[aps,prb,eqsecnum,multicol,epsf]{revtex}
\begin{document}
\input epsf
\bibliographystyle{prsty}
\draft
\preprint{NSF-ITP}
\title{Conductance fluctuations at the integer quantum Hall plateau transition}
\author{Sora Cho}
\address{Department of Physics, University of California,
        Santa Barbara, California 93106}
\author{Matthew P.A. Fisher}
\address{Institute for Theoretical Physics,
        University of California, Santa Barbara, California 93106\\
        and Department of Physics, University of California,
        Santa Barbara, California 93106}
\date{\today}
\maketitle

\begin{abstract}
We study numerically conductance fluctuations near the
integer quantum Hall effect plateau transition.
The system is presumed to be in a mesoscopic regime, with
phase coherence length comparable to the system size. 
We focus on a two-terminal conductance $G$ for square samples, considering both
periodic and open boundary conditions transverse to the current.
At the plateau transition, $G$ is broadly distributed,
with a distribution function close to uniform on the interval
between zero and one in units of $e^2/h$.
Our results are consistent with a recent experiment by
Cobden and Kogan on a mesoscopic
quantum Hall effect sample.
\end{abstract}

\begin{multicols}{2}

\section{Introduction}

One of the early surprises of the burgeoning field
of mesoscopic physics some ten years ago was the observation
of large conductance fluctuations in small metallic samples.\cite{ssp91,mps91}
Metals with phase coherence lengths exceeding their size were found
to exhibit sample (or field) specific fluctuations in their conductance.
The magnitude of the fluctuations -- of order $e^2/h$ --
was essentially independent
of the mean conductance, leading to the name 
``universal conductance fluctuations."  Theoretical explanations are based on
models of diffusing electrons, in which localization effects can be
ignored.\cite{palee87}  Generally, this requires that the mean conductance
is much larger than $e^2/h$, a condition fulfilled in the
experiments.  

One of the striking features of the plateau transitions in the quantum
Hall effect,\cite{qhe90,huckestein95} is that the magnitude of the macroscopic
longitudinal conductivity $\sigma_{xx}$ is both metallic -- independent of
temperature as $T \rightarrow 0$ -- and of order $e^2/h$.
Conventional localization effects are inoperative due to the strong
applied magnetic field.   
For the transition from insulator to the first filled Landau level,
the experimental values\cite{shahar95} 
for the conductivity tensor are consistent with
\begin{equation}
\sigma_{xx}=\sigma_{xy}={1\over 2}{e^2\over h}.
\end{equation} 
These macroscopic conductivities are self-averaged,
since the sample sizes are much bigger than the phase coherence length.
Several authors have given theoretical arguments
in support of these values,\cite{huckestein95,huo93,kivelson92,chklovskii93}
although it is unclear that the averaging process is appropriate 
to the experiment.

Recently Cobden and Kogan have measured the conductance
of a small quantum Hall effect sample, in the mesoscopic regime.\cite{cobden96}
They find large fluctuations in a two-terminal conductance near the plateau
transitions, as they vary the carrier density with a gate voltage.
Specifically, the conductance seems to be almost uniformly distributed
on the interval between zero and one in units of $e^2/h$.  In striking contrast
to conventional metallic samples, the magnitude of the
conductance fluctuations is comparable to the mean conductance,
$\overline{G} \approx (1/2) e^2/h$.

In this paper we compute the conductance fluctuations employing
a simple network model of the inter=ger quantum Hall effect (IQHE)
 plateau transition.\cite{wang96,fastenrath92}  
We extract a
mesoscopic two-terminal conductance and its sample (and field) specific
fluctuations.  Our results are entirely consistent with the 
Cobden and Kogan experiment.  Right at the transition,
the conductance distribution function is essentially uniform
on the interval from zero to one in units of $e^2/h$.  

Our paper is organized as follows.  In Sec. II we briefly review
the network model, specifying the appropriate
geometry and boundary conditions.
The results for the conductance and its distribution are presented
in Sec. III.  Sec. IV is devoted to a brief discussion.

\section{the network model}

To model the IQHE
plateau transition, we employ Chalker and Coddington's 
network model.\cite{chalker88}
In this model the interactions between the electrons are ignored.
In their original formulation,
the impurity potential was assumed to be slowly varying on the scale
of the magnetic length.  The semiclassical trajectories
moving along equipotentials
were modeled via ballistic chiral propagation along
the links of a network.  Quantum tunneling at saddle points
between nearby equipotentials was incorporated
via tunneling at node parameters, connecting two incoming and
two outgoing links. For simplicity, the nodes and links
were placed on a regular (square) lattice.
Randomness was incorporated via phase
factors for propagation along the links,
which were assumed to be independent and
uniformly distributed between $0$ and $2 \pi$.

In this paper we focus on the behavior near the
plateau transition.  Being a continuous (second order) phase
transition, we expect that universal critical properties
(including conductance fluctuations) should not depend on details
of the model.  Thus, for example, the results obtained should also
apply to systems for which the potential is {\it not}
varying slowly on the scale of the magnetic length.
Extensive numerical simulations
that have extracted the critical exponent $\nu$ for the
diverging localization length support this supposition.\cite{huckestein95}
For example, Lee, Wang, and Kivelson\cite{dhlee93} have shown that inclusion of
random scattering at the nodes gives the same value for $\nu$
as in Chalker's original random-phase model.\cite{chalker88}
Moreover, consistent estimates for $\nu$ have also been
obtained from other numerical approaches, such as
Thouless number studies of lowest Landau level Hamiltonians.
\cite{huckestein95,fastenrath92}
A more serious concern is the legitimacy of ignoring
Coulomb interactions between the electrons.  It is conceivable
that interactions -- particularly long-ranged Coulomb forces -- 
might be a ``relevant" perturbation at the noninteracting
phase transition (fixed point), leading to new critical properties.
\cite{jungwirth96,dhlee96}
However, experimental measurements of $\nu$ seem to coincide
with the non-interacting value.  Being practical, we ignore
Coulomb interactions, and adopt Chalker and Coddington's network
model.\cite{chalker88}

To be specific, we study a square lattice network of nodes and links,
as depicted in Fig. 1.  The network is connected
to two leads -- to the right and left.  
The distance between the leads, measured in units
of the network lattice spacing, is denoted by $L_x$.
Of interest is the two-terminal
conductance between these two leads.  
The sample has a width, denoted $L_y$, in the transverse
direction.
We consider two boundary conditions in the transverse direction:
(i) periodic boundary conditions and (ii) open boundary
conditions.  The case of open boundary conditions
corresponds closely to the
experimental geometry of Cobden and Kogan.\cite{cobden96}
In this case, edge states dominate the transport in the
IQHE plateau.

Quantum tunneling at each node is
represented by a 2 by 2 matrix,
\[  \left( \begin{array}{c} w_{\rm out} \\
w_{\rm in} \end{array} \right) =
 \left( \begin{array}{ccc}
    \cosh\theta & & \sinh\theta \\ \sinh\theta & & \cosh\theta
\end{array} \right)
\left( \begin{array}{c} v_{\rm in} \\ v_{\rm out} \end{array} \right), \]
where $v$ and $w$ represent complex amplitudes
for incoming and outgoing electron waves to the right and left, 
respectively, of a given node (see Fig. 1).
By construction, this matrix conserves the current,
$|w_{\rm in}|^2+|v_{\rm in}|^2 =|w_{\rm out}|^2+|v_{\rm out}|^2$.
The node parameter $\theta$ determines the degree of backscattering
at the node.  So, for example,
an incident wave, say $w_{\rm in}$, is back-scattered into $w_{\rm out}$
with probability $\tanh^2(\theta)$.  
To make the model invariant under a $\pi/2$ spatial
rotation, the node parameters take two values,
$\theta_1$ and $\theta_2$,
in alternating columns, and are chosen to satisfy
the condition $\sinh(\theta_1)\sinh(\theta_2)=1$.
Randomness is incorporated via random phase factors
along links.

 From symmetry, the plateau transition occurs when
$\tanh(\theta_1)=\tanh(\theta_2)$, or $\theta=\theta_c=\ln(1+\sqrt{2})$.
It is thus convenient to define a variable 
$\Delta$, which measures the ``distance'' 
to the transition
\begin{equation}
\Delta=\tanh(\theta_{1})-\tanh(\theta_2).
\end{equation}
This parameter lies in the range
$-1\leq\Delta\leq 1$, and vanishes right at the plateau transition.

\begin{figure}
\epsfxsize=2.7in
\centerline{\epsfbox{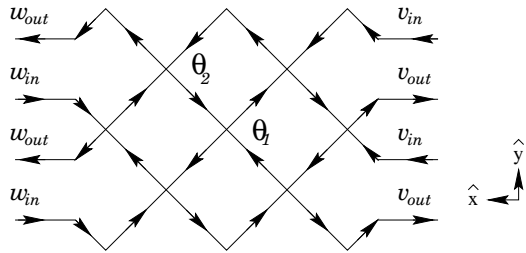}}
\begin{minipage}[t]{8.1cm}
\caption{Schematic representation of the network model
for a square sample with size $L_x=L_y=4$.  
The arrows indicate the direction of 
wave propagation along the links, and $\theta_1$
and $\theta_2$ specify 
scattering at the nodes.  The two-terminal conductance $G$ is measured
between the right and left leads.}
\end{minipage}
\end{figure}

For open boundary conditions in the transverse $(y)$ direction,
the nodes on the top and bottom edges are modified to be
\[  \left( \begin{array}{c} w_{\rm out} \\
w_{\rm in} \end{array} \right) =
 \left( \begin{array}{cc}
    1 & 0 \\ 0 & 1
\end{array} \right)
\left( \begin{array}{c} v_{\rm in} \\ v_{\rm out} \end{array} \right) \]
so that $w_{\rm out}=v_{\rm in}$ and $w_{\rm in}=v_{\rm out}$.
In this case, the boundary breaks (lowest Landau level) particle-hole
symmetry, $\Delta \rightarrow - \Delta$, just as the edges do in
a real physical system.  In the Hall plateau phase,
$\Delta > 0$, extended edge states confined to the top and bottom
boundaries of the sample are expected.  In the localized insulator,
corresponding to $\Delta < 0$, all states are localized, even near
a boundary.  In real systems, the presence of an edge state
accounts naturally for the quantized Hall conductivity.

With transverse periodic boundary conditions,  
all nodes (in each column) are identical.
In this case, there are of course no edge states.
The two-terminal conductance is only sensitive to
extended bulk states, present at the plateau transition.
The finite system is particle-hole symmetric
($\Delta \rightarrow -\Delta$), as will be clear from the numerical results.

We compute numerically the total transfer matrix $T$ 
in the x direction,
for a network of width $L_y$ and length $L_x$.
This $L_y$ by $L_y$  matrix relates the incoming and outgoing amplitudes
in one lead -- the $v$'s -- to the amplitudes in the other lead - 
the $w$'s, and can be written schematically as $W=TV$,
where $W$ and $V$ are vectors with $L_y$ elements.
To extract the two-terminal conductance $G$,
it is useful to write $T$ in the form
\[  \left( \begin{array}{c} W_{\rm out} \\
W_{\rm in} \end{array} \right) =
 \left( \begin{array}{ccc}
    A & & B \\ C & & D
\end{array} \right)
\left( \begin{array}{c} V_{\rm in} \\ V_{\rm out} \end{array} \right), \]
where $V_{\rm in}$
denotes an $L_y/2$ column vector of the incoming
amplitudes in the right lead, $V_{out}$ the outgoing,
and similarly for the other lead.
This can be inverted to obtain the 
$S$ matrix relating all the incoming to outgoing modes, as
\[  \left( \begin{array}{c} W_{\rm out} \\
V_{\rm out} \end{array} \right) =
 \left( \begin{array}{lll}
    B D^{-1} & & A-B D^{-1} C \\D^{-1}  & & -D^{-1} C
\end{array} \right)
\left( \begin{array}{c} W_{\rm in} \\ V_{\rm in} \end{array} \right). \]
The two-terminal conductance can be expressed in terms of the
$L_y/2$ by $L_y/2$ transmission matrix $t$,
which relates the incoming amplitudes of one lead
to the outgoing amplitudes in the other lead:
$V_{\rm out} = t W_{\rm in}$.
Thus we have $t=D^{-1}$.
The two-terminal conductance follows readily from $t$ as,\cite{dfisher81}
\begin{equation}
G={e^2\over h}{\rm tr}[t t^+] .
\end{equation}

Below we focus on the two-terminal conductance for square samples
of size $L=L_x=L_y$=4, 8, 16, and $24$.  
The distribution function is obtained by
evaluating $G$ for many different samples.  
We typically take 
$5\times 10^4$ samples.

\section{results}

The mean conductance, denoted $\overline{G}$, is obtained by averaging $G$
over a large number of samples.  In Fig. 2(a), $\overline{G}$ is plotted versus
the control parameter $\Delta$, for four different sample sizes,
all with periodic boundary conditions in the transverse direction.
(In this and later figures, $G$ is plotted in units $e^2/h$.)
As expected the mean conductance is largest at the plateau transition,
$\Delta=0$, reflecting transport via bulk delocalized states, and
falls off towards zero away from the transition.
For the larger system sizes, the peak in $\overline{G}$ narrows.
With open boundary conditions, the mean conductance rises from zero to one,
as shown in Fig. 2(b).  As the sample size increases, the ``step"
becomes less and less rounded.  Together, the two sets of data for $\overline{G}$,
resemble experimental plots of the
macroscopic conductivities, $\sigma_{xx}$ and $\sigma_{xy}$,
when plotted versus electron density.  However, a direct comparison
with macroscopic conductivities is delicate, since the ensemble averaging
procedure that we are using may {\it not} be appropriate
to the thermal averaging taking place in macroscopic experimental samples.

 From scaling arguments, one expects that the mean conductances,
$\overline{G}(L,\Delta)$, should be only a function of the
single scaling parameter,\cite{huckestein95} 
$L^{1/\nu}\Delta$ for large enough $L$
and small enough $\Delta$.  Here $\nu$ is the localization length
critical exponent, which has a value $\nu \approx 7/3$.
In Fig. 3, the the data for $\overline{G}$ in Fig. 2, are replotted
versus the scaling variable $L^{1/\nu}\Delta$, with $\nu=7/3$.
Although the smaller sizes show marked finite-size scaling corrections,
the data collapse is satisfactory at the larger sizes.

It is amusing that with periodic boundary conditions,
$\overline{G}$ at the plateau transition $\Delta=0$,
is very close to $1/2$ for the largest sizes,
the same value as the experimentally
measured {\it macroscopic} $\sigma_{xx}$.
But this is probably coincidental, since our averaging procedure is
not appropriate for macroscopic samples.
With open boundary conditions, the mean conductance
is not invariant under $\Delta \rightarrow -\Delta$, due
to the breaking of particle-hole symmetry by the boundaries.
At the plateau transition, $\overline{G}$ is slightly larger than $1/2$,
roughly $0.65$ for the larger sizes.  The increase of
$\overline{G}(\Delta=0)$ when changing the boundary conditions from
periodic to open can presumably be attributed
to an additional contribution coming from 
edge currents.
 
\begin{figure}
\epsfxsize=3.5in
\centerline{\epsfbox{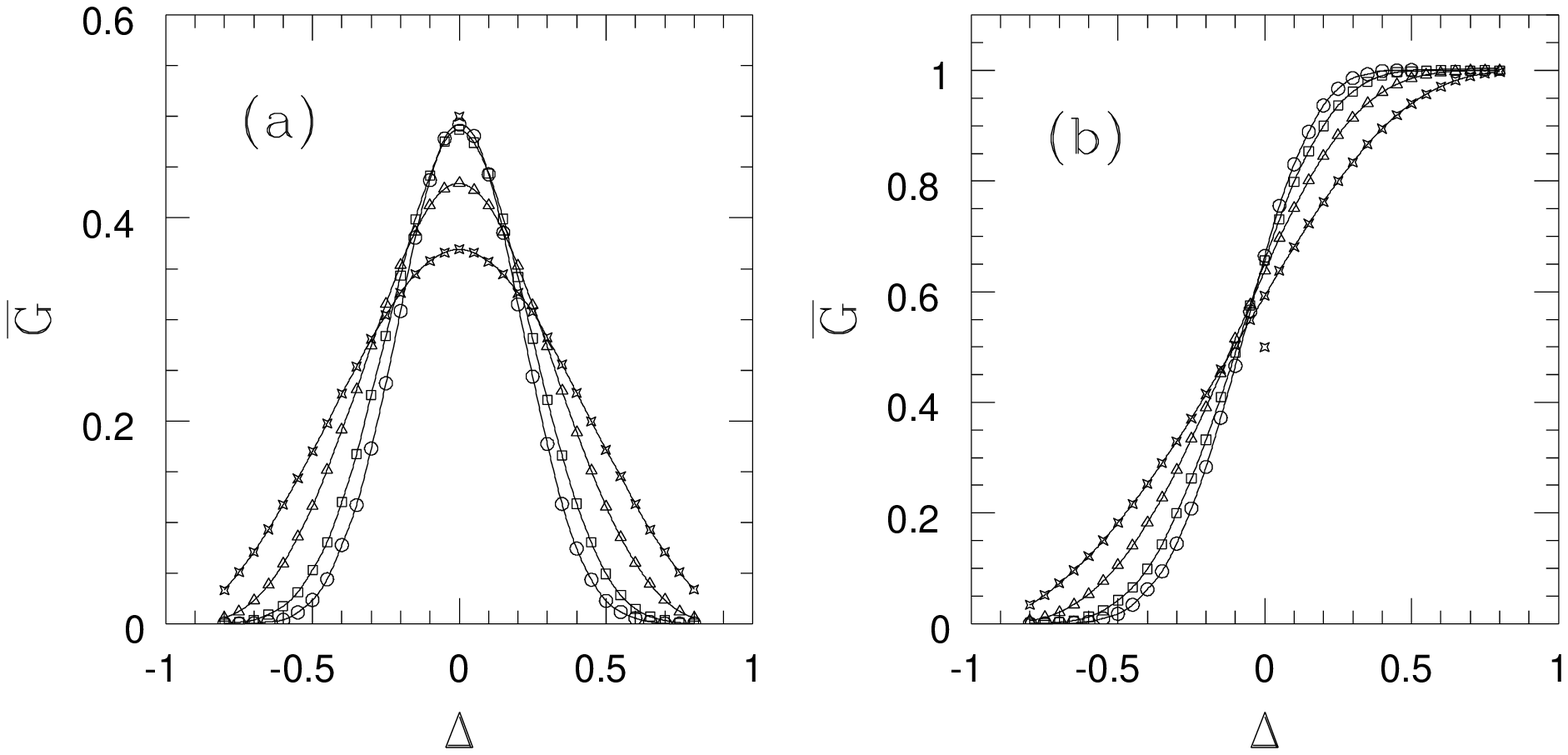}}
\begin{minipage}[t]{8.1cm}
\caption{ The mean conductance ${\overline G}$ plotted vs
 $\Delta$ for square systems of four different sizes
 $L=4(\times)$, $L=8(\triangle)$, $L=16(\Box)$, and $L=24(\bigcirc)$
with (a) periodic and (b) open boundary conditions.}
\end{minipage}

\epsfxsize=3.5in
\centerline{\epsfbox{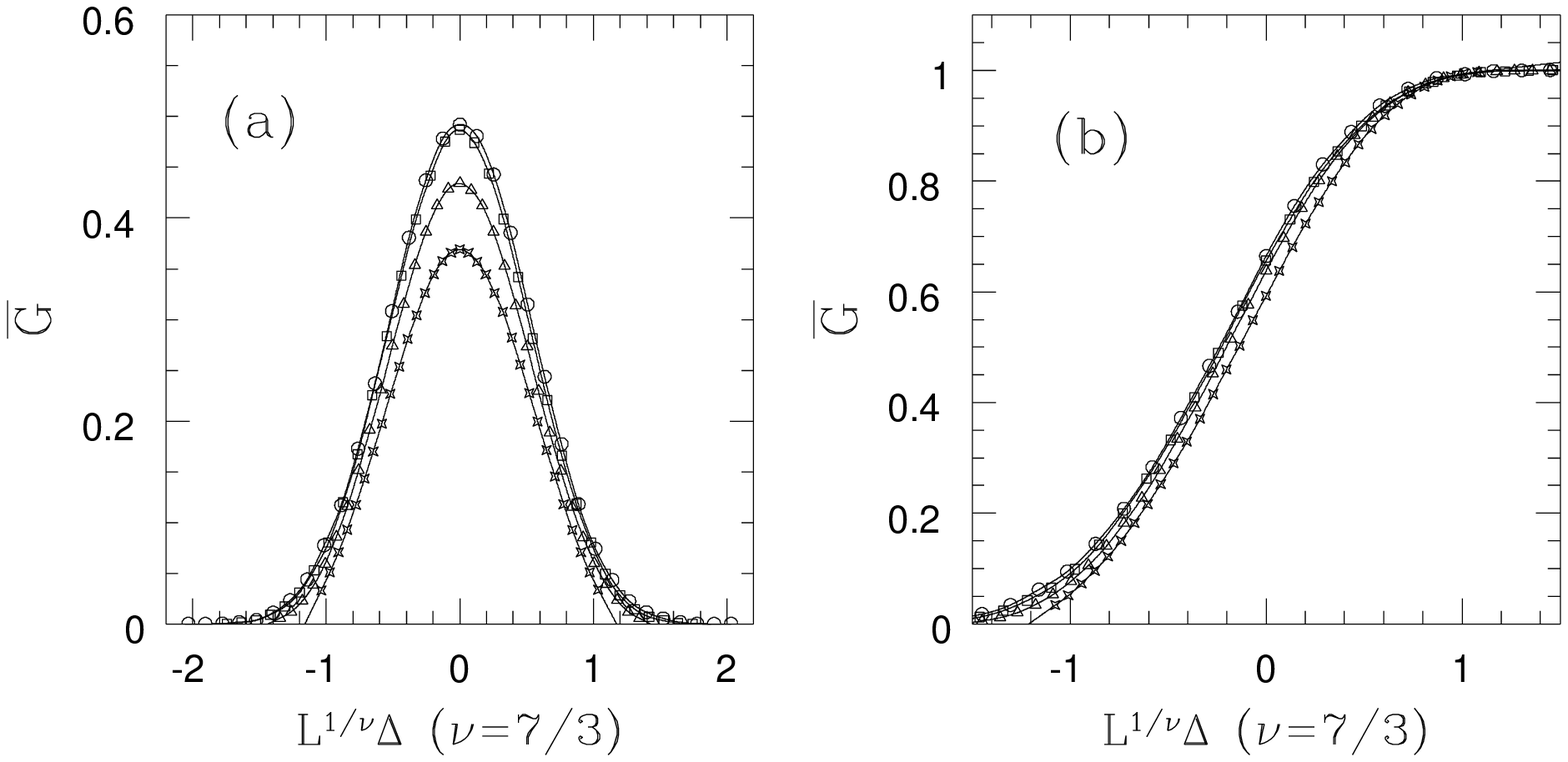}}
\begin{minipage}[t]{8.1cm}
\caption{  Scaling collapse of the mean conductance from Fig. 2,
plotted vs the parameter $\Delta L^{1/\nu}$
with $\nu=7/3$, for (a) periodic and (b) open boundary conditions.}
\end{minipage}
\end{figure}

To characterize the conductance fluctuations first consider
the root-mean-square conductance,
defined as 
\begin{equation}
\delta G= \sqrt{\overline{G^2} - \overline{G}^2},
\end{equation} 
where the
overbar denotes an ensemble average over different samples.
In Fig. 4, we plot $\delta G$ versus $\Delta$
for four different sample sizes, with periodic boundary conditions
in (a) and open in (b).  For both boundary conditions,
there are large fluctuations near the plateau transition,
with $\delta G(\Delta =0) \approx 0.3$.  Away from the transition, the fluctuations drop off
rapidly with increasing system size, as expected.
In both cases, the peaks sharpen
with increasing $L$, as expected from finite size scaling.
The physical origin of the slight double-peaked structure in Fig. 4(a)
is unclear.  In contrast to the mean conductance itself,
the root-mean-square conductance is relatively insensitive to finite-size 
effects at the transition.

\begin{figure}
\epsfxsize=3.5in
\centerline{\epsfbox{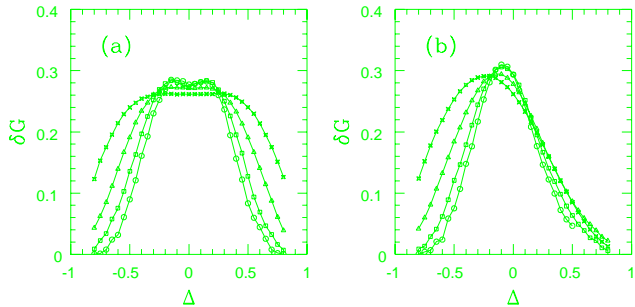}}
\begin{minipage}[t]{8.1cm}
\caption{  Root mean square conductance $\delta G$ plotted vs
 $\Delta$ for a square system of four different sizes:
$L=4(\times)$, $L=8(\triangle)$, $L=16(\Box)$, and $L=24(\bigcirc)$
with (a) periodic and (b) open boundary conditions.}
\end{minipage}
\end{figure}

More informative than the root-mean-square conductance,
is the full conductance distribution function,
denoted $P(G)$.  To obtain this, we simply make a histogram plot of 
the number of samples with conductance $G$, for a very large ensemble.
In Fig. 5, we plot the conductance distribution function
at the plateau transition ($\Delta=0$), for the largest system size
$L=24$, with (a) periodic and (b) open boundary conditions.
In both cases, the conductance is very broadly distributed,
roughly uniform over the interval from zero to one $e^2/h$.
Above $G=e^2/h$, the distribution functions drop off rapidly,
although there is a slightly larger ``tail" with open boundary
conditions, presumably due to edge current contributions.

\begin{figure}
\epsfxsize=3.5in
\centerline{\epsfbox{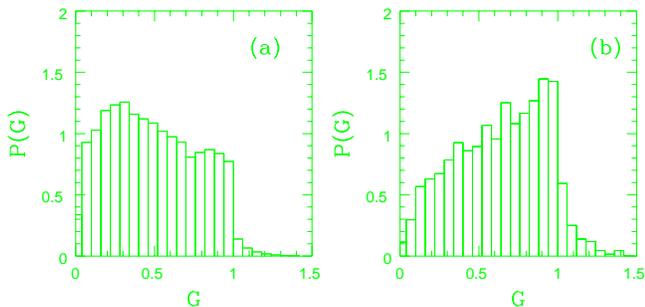}}
\begin{minipage}[t]{8.1cm}
\caption{  Conductance distribution function right at the plateau
transition, $\Delta=0$, for the system size $L=24$ 
with (a) periodic and (b) open boundary conditions.
The distribution functions are normalized to unity.}
\end{minipage}
\end{figure}

\section{Discussion}
 
The conductance distribution functions obtained numerically
compare favorably with those measured
in the Cobden and Kogan experiment on 
a mesoscopic Hall system.\cite{cobden96}  
In this experiment, the two-terminal conductance of a small
sample ($0.6\times 0.6$ $\mu m^2$) was measured as a function
of carrier density, by varying a gate potential $V_g$.
Large fluctuations in the conductance were seen upon varying 
$V_g$ through the plateau transitions, as shown in Fig. 6(a).  
In the Hall plateaus themselves,
smaller fluctuations were observed.
This behavior is consistent with our numerics for the root-mean-square 
fluctuations, $\delta G$ in Fig. 4, which are largest
at the plateau transition. 
 From the data with gate voltages near the
plateau transition, 
Cobden and Kogan obtained a conductance distribution function,
shown in Fig. 6(b), 
which is roughly uniform on the interval from zero to one.
Under the ``ergodic" assumption that varying the gate potential
is equivalent to changing the impurity configuration (i.e., the sample),
we can compare their distribution function with ours, which was 
obtained by taking an ensemble average {\it at} the plateau
transition (see Fig. 5).  
The similarity is striking.

\begin{figure}
\epsfxsize=3.5in
\centerline{\epsfbox{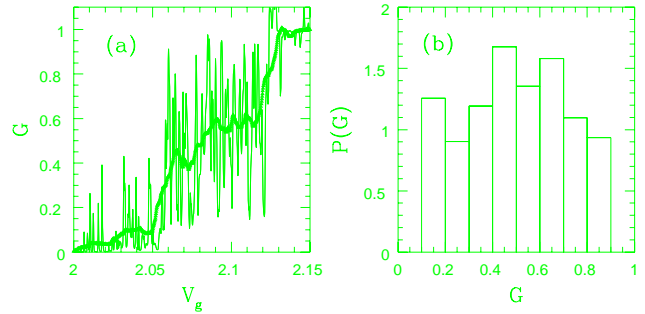}}
\begin{minipage}[t]{8.1cm}
\caption{Cobden and Kogan's experimental data:
(a) Two-terminal conductance plotted vs
the gate voltage $V_g$.  The thick solid line
is the same data averaged over a $V_g$ interval of 16 mV.  
(b) Conductance distribution of data points near the
plateau transition, in the interval $2.06\leq V_g\leq 2.12$.}
\end{minipage}
\end{figure}

To more closely mimic the experimental procedure, we have computed
the conductance for a {\it given} sample,
as a function of $\Delta$.
This is shown in Fig. 7, for a square system of size
$L=24$ with both periodic and open boundary conditions.
Notice the very large fluctuations in the transition region.
The behavior in Fig. 7b with open boundary conditions is very similar to the
``raw" experimental data of conductance versus gate potential. 

In addition to extracting conductance fluctuations for square samples,
we have studied systems with various different aspect rations, $L_x/L_y$.
For aspect rations between roughly $1/3$ and $3$, the 
qualitative results are essentially unmodified.  For very long sample,
however, 
$L_x >> L_y$ we start seeing effects of one-dimensional
localization.

\begin{figure}
\epsfxsize=3.5in
\centerline{\epsfbox{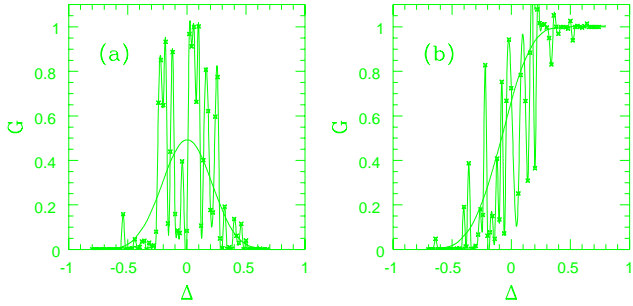}}
\begin{minipage}[t]{8.1cm}
\caption{ Conductance plotted vs $\Delta$ of
a single sample for the system size $L=24$ with
(a) periodic and (b) open boundary conditions.  The solid lines
are the mean conductance, which was obtained by taking an ensemble average.}
\end{minipage}
\end{figure}

In summary, it appears that the quantum Hall plateau transition
provides an ideal arena for studying
finite-size effects on random phase transitions.
The agreement between simple models of non-interacting electrons,
and the experimental data is striking.  Among the open issues
is the role of Coulomb interactions, which have been ignored
in the numerics.  
Will they change the critical behavior, and possibly
modify the conductance fluctuations?
Moreover, even without interactions, 
an analytic description of the transition is lacking, for either
fluctuations or average properties.  One can only hope that
the relative experimental accessibility of
the quantum Hall plateau transition, will spur further
theoretical developments.

\section*{Acknowledgments}

We thank David H. Cobden for generously sharing his experimental data.
It is a pleasure to acknowledge Nathan Argaman for fruitful conversations.
We are grateful to the National Science Foundation for support under Grants
No. PHY94--07194, No. DMR--9400142, and No. DMR-9528578.

\bibliography{/usr/home/spock/sora/paper/biblio/ref}

\end{multicols}
\end{document}